\documentstyle[12pt,multicol,amsfonts,epsf]{article}

\newcommand{\beq}{\begin{equation}}
\newcommand{\eeq}{\end{equation}}
\newcommand{\beqy}{\begin{eqnarray}}
\newcommand{\eeqy}{\end{eqnarray}}

\newtheorem{Definition}{Definition}

\newtheorem{Theorem}{Theorem}

\newenvironment{Proof}{{\it Proof: \,}}{$\Box$ \vspace{0.3cm}}

\newenvironment{Definition*}{{\bf Definition}}{}

\makeatletter
\def\@beginTheorem#1#2{\trivlist \item[\hskip \labelsep{\bf #1\ #2}]}
\def\@opargbegintheorem#1#2#3{ \trivlist
      \item[\hskip \labelsep{\bf #1\ #2\ (#3)}]}
\makeatother

\makeatletter
\def\@beginLemma#1#2{\trivlist \item[\hskip \labelsep{\bf #1\ #2}]}
\def\@opargbeginLemma#1#2#3{ \trivlist
      \item[\hski
 Hence we have the same statements
about the increase of the supports for increasing depth, where
the local transformations are not counted for the depth.
p \labelsep{\bf #1\ #2\ (#3)}]}
\makeatother

\makeatletter
\def\@beginDefinition#1#2{\trivlist \item[\hskip \labelsep{\bf #1\ #2}]}
\def\@opargbeginDefinition#1#2#3{ \trivlist
      \item[\hskip \labelsep{\bf #1\ #2\ (#3)}]}
\makeatother

\makeatletter
\def\@beginCorollary#1#2{\trivlist \item[\hskip \labelsep{\bf #1\ #2}]}
\def\@opargbeginCorollary#1#2#3{ \trivlist
      \item[\hskip \labelsep{\bf #1\ #2\ (#3)}]}
\makeatother

\makeatletter
\def\@beginExample#1#2{\trivlist \item[\hskip \labelsep{\bf #1\ #2}]}
\def\@opargbeginExample#1#2#3{ \trivlist
      \item[\hskip \labelsep{\bf #1\ #2\ (#3)}]}
\makeatother

\def\Z{{\mathbb{Z}}}
\def\R{{\mathbb{R}}}

\newcommand{\cH}{{\cal H}}

\title{Quantum thermodynamics with missing reference frames:
Decompositions of free energy into non-increasing components}

\author{ Dominik
Janzing\thanks{e-mail: 
janzing@ira.uka.de} 
\\ \small Institut f{\"u}r Algorithmen und Kognitive Systeme,
Universit{\"a}t Karlsruhe,\\[-1ex] \small Am Fasanengarten 5,
D-76\,131 Karlsruhe, Germany}

\begin{document}
\maketitle

\begin{abstract}  
If an absolute reference frame with respect to time, position, or 
orientation
is missing one can only implement
quantum operations which are covariant with respect 
to the corresponding unitary symmetry group $G$. 
Extending observations of Vaccaro et al., I argue that
the free energy of a quantum system with $G$-invariant Hamiltonian
then splits up into the Holevo information
of the orbit of the state under 
the action of $G$  and the free energy of its orbit average.
These two kinds of free energy cannot be converted into each other. 
The first component is subadditive and the second superadditive; 
in the limit of infinitely many copies only the usual
free energy matters. 

Refined splittings of free energy 
into more than two independent (non-increasing) terms can be defined by averaging
over probability measures on $G$ that differ from the Haar measure. 

Even in the presence of a reference frame,
these results provide lower bounds  on the amount of 
free energy that is lost after applying 
a covariant channel. 
If  the channel 
properly decreases 
one  of these quantities, it decreases the free energy 
necessarily at least by the same amount, since
it is unable to convert
the different forms of free energies into each other. 
\end{abstract}

\section{Introduction}

Free energy is among the most important concepts of thermodynamics
since it formalizes the fact that the 
usability of energy resources depends also on their
{\it entropy}. 
Roughly speaking, the idea is that
in an energy conversion process the target system should typically be 
provided with energy without transferring entropy 
(like increasing the kinetic of potential 
energy or a macroscopic body, for instance). 
Therefore, the worth of a system for
being an energy source depends on the question to what extent one can
extract energy from it without releasing too much entropy $S$ since 
the transfer of the dispensable entropy to the environment requires the
additional amount $S k_B\,T$ of energy (here $k_B$ is Boltzmann's constant 
and $T$ is the  temperature of the heat reservoir where the entropy is
transferred to, e.g., the environment).
Hence the 
amount of 
work that can be  extracted from a physical system 
is not given by its inner energy. Instead, it depends also 
on the entropy and on a fixed reference
temperature, namely the temperature of the environment which is potentially used
as an entropy sink. Conversely, a system 
that has no inner energy at all (like a degenerate two-level system)
 can be used to extract energy from the environment
 if its physical state 
is not the maximal entropy state. In other words, 
{\it information} can directly be used to  extract work from the surrounding heat bath \cite{Szilard,JWZGB,Ja00}. This fact implies on the other hand
that the initialization of
bits requires energy resources, an observation which is usually
referred to as Landauer's principle (cp. \cite{Landauer:61,BennettThermoReview}). 
All these statements can be brought into a consistent picture
by the notion of free energy: for instance,  the initialization 
of a non-degenerate two-level system to a well defined pure state
increases its free energy  and requires therefore resources of free
energy. 

For a quantum system with  
density operators $\rho$ and Hamiltonian $H$ one can define 
(in analogy to the classical definition, see \cite{Callen})
the free
energy by the difference 
\begin{equation}\label{Free1}
tr(\rho H) -S(\rho)\, k_B T\,,
\end{equation}
where 
\[
S(\rho):= -tr(\rho \ln \rho)
\]
is the von-Neumann entropy\footnote{Note that 
it is not straightforward to replace the classical quantity $S$ with 
von-Neumann entropy
in quantum mechanics \cite{AllahverdyanGibbs}. 
However, it goes beyond 
the scope of this article to discuss this issue.
We assume here that the free
energy in eq.~(\ref{Free1}) is nevertheless a reasonable quantity
for quantum thermodynamics.}
of $\rho$. 
Since the Gibbs state
\begin{equation}\label{Gibbs}
\gamma_T:=\frac{e^{-H/(k_B T)}}{tr\Big(e^{-H/(k_B T)}\Big)}
\end{equation}
minimizes (\ref{Free1}) it is convenient to redefine free energy by
\begin{equation}\label{FreeTotal}
F(\rho):= tr(\rho H) -S(\rho) \,k_B T -
tr(\gamma_T H) +S(\gamma_T)\, k_B T\,,
\end{equation}
implying that $F(\rho)=0$ if and only if $\rho=\gamma_T$. 

It is known that no physical process which uses no additional energy resources
can
convert $\rho$ into a state $\tilde{\rho}$ with 
$F(\tilde{\rho})> F(\rho)$. This follows already from the fact that
$F(\rho)$ is up to the constant
$k_B T$ the Kullback-Leibler distance between $\rho$ and $\gamma$ 
\cite{OhyaPetz} and that it is not possible to create
non-equilibrium states from $\gamma$ without using additional energy 
resources. Therefore $\gamma$ is invariant 
with respect to all those operations and to increase $F$ would mean to 
increase the distance to $\gamma$ in contradiction to the
fact that no operation can increase Kullback-Leibler distances between
density operators
\cite{OhyaPetz}. 

Even though the monotonicity of 
free energy is maybe the most important constraint on the
possible operations
one should not forget that additional constraints arise in particular for quantum systems. However, they depend on 
additional assumptions on the set of physical processes.
Refs.~\cite{Ja00,JWZGB,AllahverdyanNano,Allahverdyan} 
consider work extraction from quantum systems
by unitary operations. In these models, the amount of extractable work
does not only depend on the inner energy and the entropy of the system. 
Instead,
it can only be calculated from more detailed information on the spectrum
of the density operator. 

In \cite{clock} we have furthermore  considered {\it timing information} 
as a kind
of thermodynamic resource. The idea was the following. Defining a physical 
system that includes all available clocks, no operation on the joint system
can increase the information about an externally defined time reference
frame. Every quantum state that is prepared in a superposition of different
energy eigenstates
with  well-defined phase with respect to the external reference
frame, provides some information on the latter. The impossibility 
to create information  on the external time can also be interpreted
as the impossibility to prepare superpositions of different energy eigenstates
with well-defined phase. 
Such kind of coherent 
superpositions are therefore a special kind of deviation from equilibrium
which could be considered as a resource in its own right. 
Whereas 
the complexity for communicating reference frames 
for time,  position, and orientation has been extensively studied
in the literature (e.g. \cite{RefEfficient,GroverRudolph}) 
and its cryptographic power has been pointed out \cite{Decoherence-full},
the thermodynamical relevance
of reference frames has mainly be considered in 
 \cite{SynchrEntropy} and \cite{GroupCovariantThermo}.
Whereas the first article considers the thermodynamic cost of establishing
reference frames with classical communication,
the authors in \cite{GroupCovariantThermo} 
observe that the worth of thermodynamic resources
is reduced by missing reference frames. The latter observation
is the venue of this article. 

The paper is organized as follows.
In Section  \ref{Sec:Cov} we sketch the idea
to restrict the set of operations to covariant maps, i.e.,
those that can be implemented without refering to an external frame.  
We rephrase the idea of Vaccaro et al.~\cite{GroupCovariantThermo}
to consider a  thermodynamic theory 
that is modified by the additional constraint
of covariance. 
 
In Section \ref{Sec:Split} we will show explicitly that 
the covariance condition implies a splitting of free energy into
two terms which cannot be converted into each other. 
We refine this splitting of free energy into arbitrarily many
terms reflecting the fact that different kind of 
timing information that refer
to different {\it time scales} cannot be converted into each other. 
The theory can be generalized to other covariance conditions that may stem, for instance, from missing spatial or 
rotational reference frames provided that the considered unitary
symmetry operation
commutes with the Hamiltonian. 
In Section \ref{Sec:Super} we argue why the splitting loses its relevance in the 
macroscopic limit of a large number of identical systems.
In Section \ref{Sec:Appl} we will show that the results have implications
also for situations where reference frames are available. This is because
every covariant operation that decreases one kind of free energy decreases
also the total amount of free energy and this loss is clearly 
irreversible even if a reference frame is available. 
We sketch how to apply  this idea to
time-invariant passive devices (i.e. devices without energy source) 
in optics or in electrical engineering. Then our results 
imply that a device that causes
an indeterministic time delay of the output signal 
causes necessarily a loss of
free energy.

\section{Covariant operations}

\label{Sec:Cov}

The set of possible operations on a quantum system is given by the set of
completely positive trace-preserving maps \cite{NC}. 
In \cite{clock} we have argued that not every CP map $C$ can be implemented
if no time reference frame is available. 
Consider a quantum system with free evolution
\[
\alpha_t(\rho)=e^{-iHt} \,\rho\,e^{iHt}\,.
\]
Assume that the state $\rho$ was prepared at time $t=0$. 
If a person implements an operation $C$ at time instant $s$ the state
of the system after $C$ was implemented  is described by
\[
C(\alpha_s(\rho))\,.
\]
Looking at the system later it is described by the state
\begin{equation}\label{Rtime}
\alpha_{t-s}\Big(C(\alpha_s(\rho))\Big)\,,
\end{equation}
if $t$ is the time that has  passed since the system was prepared. If no
clock was available during the implementation of $C$, it is  
implemented at a random time instant $s$. We may therefore define
a CP map $\overline{C}$ that results from averaging (\ref{Rtime})
over all $0\leq s \leq t$. If the quantum system has discrete spectrum and
$t$ is large compared to the time scale given by the 
inverse of the minimal distance between its energy eigenvalues, 
$\overline{C}$ satisfies approximatively
the covariance condition
\begin{equation}\label{Cov}
\alpha_t \circ \overline{C}=\overline{C} \circ \alpha_t,\,\,\,\,\,\forall t\,.
\end{equation}
In \cite{clock} we have therefore assumed that the set of operations
which can be performed without additional clock is given by those that
satisfy the above covariance condition (\ref{Cov}). In \cite{TimeCovariant}
we have analyzed this class  of CP maps in full detail.

With the same arguments one can restrict the set of available operations
to those satisfying covariance conditions with respect to other symmetry 
groups if the corresponding reference frame is not available\footnote{For a model
for a formulation of 
quantum mechanics that avoids absolute reference frames see \cite{PoulinRelation}.}. 
Interesting instances are given by the group of space translations or by the rotation 
symmetry \cite{GroupCovariantThermo}.
The symmetry is represented 
by a unitary group $U_g, g \in G$ acting on the Hilbert space of
the considered system. We assume that
$[U_g, H]=0$, otherwise $(U_g)$ would not be a symmetry group of the 
Hamiltonian and the missing reference frame would even make the definition
of $H$ impossible. 
Every transformation is 
covariant with respect to $G$, i.e., we can only implement
a CP map $C$ with 
\begin{equation}\label{CovGen}
C(U_g \rho U_g^\dagger)=U_g C(\rho) U_g^\dagger\,.
\end{equation}
 
It has already been observed in Ref.~\cite{GroupCovariantThermo} 
that the absence
of a reference frame puts thermodynamically relevant 
constraints on the set of available 
operations. The idea is that the 
system may contain
some information that is not accessible without using the frame.  
The 
authors assume that the work extractable  
from a $d$-dimensional system being in the mixed state $\rho$
is usually (if a reference frame is available) given by
\begin{equation}\label{Free}
k_BT\,\Big( \ln d - S(\rho)\Big)\,.
\end{equation}
Note that this definition of extractable work 
refers actually to thermodynamics in degenerate systems
or the infinite temperature limit, where the Hamiltonian of the system
is irrelevant and the free energy $F(\rho)$ 
is given by the difference
of the entropy to the maximally mixed state.  Within this 
thermodynamic perspective \cite{oppenheim} all maximally mixed states
are free resources whereas usual  (finite temperature and non-degenerate) 
thermodynamics
assumes all Gibbs states to be free and ``worthless'' resources.  
To consider $F$ as defined in Eq.~(\ref{FreeTotal}) 
as the extractable work is therefore
a bit more general.
We will refer to these two  points of view as the 
finite and the infinite temperature picture, respectively. 

If no reference frame is available, 
the extractable work reduces to 
\begin{equation}\label{CoFree}
W_G:=k_B T\,\Big( \ln d -S(\overline{\rho})\Big)
\end{equation}
instead of (\ref{Free}) (as the authors of \cite{GroupCovariantThermo}
observe),
where 
\[
\overline{\rho}:=\int_G U_g \,\rho\, U_g^\dagger \,d\mu(g)
\]
is the average 
of $\rho$ over the orbit of $G$, where we have implicitly
assumed for the moment 
that $G$ is a compact group and denoted its Haar measure by $\mu$.
 
Introducing the reference information by
\[
R(\rho):=S(\overline{\rho})-S(\rho)\,,
\]
the deficit between
the terms (\ref{Free}) and (\ref{CoFree}) is $R(\rho)\, k_BT$, a term called asymmetry in \cite{GroupCovariantThermo}.
The asymmetry is 
non-increasing under covariant operations since 
no trace-preserving CP map can increase
the Holevo information\footnote{This follows, for instance, if one rewrites
Holevo information as the mutual information of the bipartite state
$\sum_j p_j |j\rangle \langle j|\otimes \rho_j$. Then the statement
follows because no local operation on one system can increase
the mutual information
 \cite{LindOp} of the joint system.}. 
The authors of 
Ref.~\cite{GroupCovariantThermo} 
prove that  even covariant operations that include
measurements 
cannot increase the average asymmetry as long as the probabilities
for the measurement outcomes are $G$-invariant.
They consider a family $C_j$ of maps where each $C_j$ is a covariant CP map\footnote{Note that this class of operations does {\it not} include 
general covariant measurements
where the outcome probabilities change according to the group action.} 
and show that the average asymmetry of 
the conditional post measurement states $\rho_j:=C_j(\rho)/p_j$ with 
$p_j:=tr(C_j(\rho))$ 
cannot exceed the initial asymmetry. 
However, this generalization needs not explicitly be made
when we include a toy version of a measurement apparatus into the description. 
One can check that there exists a covariant map $C$ that transfers
the state
\begin{equation}\label{InAs}
|0\rangle\langle 0| \otimes \rho
\end{equation}
of the ``measurement apparatus'' plus system into
\begin{equation}\label{Meas}
\sum_j |j\rangle \langle j| \otimes C_j(\rho)\,,
\end{equation}
where the state $|j\rangle$ indicates that $j$ was measured.
Assuming that the group acts trivially on the ancilla system, the
asymmetry of the state (\ref{Meas}) coincides with the average
asymmetry of the ensemble $\rho_j, p_j$. 
Since we know that $C$ cannot increase
the asymmetry of (\ref{InAs}), 
 the average post-measurement 
asymmetry cannot be increased either.

In the following section we will show that the 
observations of \cite{GroupCovariantThermo} can be generalized to the
finite temperature setting and give rise to two kinds of free energy.

\section{Decomposition of free energy}

\label{Sec:Split}

The key statement of this section is that the different components
in which we decompose the free energy  are independent resources
in the sense that no covariant channel can increase them without
access to an additional energy resource. To state this formally,
we will use the notion of a passive channel:

\begin{Definition}
A trace-preserving CP map $C$ acting on a quantum system with Hamiltonian $H$ 
is called passive if $C(\gamma_T)=\gamma_T$ with the thermal state 
$\gamma_T$ 
as defined in eq.~(\ref{Gibbs}).
\end{Definition}

We have already seen 
that passivity implies $F(C(\rho))\leq F(\rho)$.
To define our decomposition of free energy, 
we assume for the moment
that the considered quantum system has discrete energy 
spectrum such that the time average
$
\overline{\rho}
$
exists. It is then given by
\[
\overline{\rho}:=\sum_j P_j \rho P_j\,, 
\]
where $(P_j)$ is the family of energy eigenprojections. 
We write the free energy $F(\rho)$ as
\[
F(\rho)=F(\rho)-F(\overline{\rho}) +F(\overline{\rho})\,,
\]
and use the fact that averaging over the time can only decrease
the free energy since the energy term in (\ref{FreeTotal}) remains the same. 
Then we have
\[
F(\rho) -F(\overline{\rho})= \Big(S(\overline{\rho})-S(\rho)\Big)\, k_BT =R\, k_B T\,,
\]
and conclude
\[
F(\rho)=R(\rho)\, k_B T + F(\overline{\rho})
\]
where we call $F(\overline{\rho})$ 
the {\it covariant free energy}. Note that $F(\overline{\rho})$ 
can be considered as the natural
 generalization of the accessible work
$W_G$ in eq.~(\ref{CoFree}) to our finite temperature setting.
To see that $F(\overline{\rho})$ is non-increasing when applying passive channels we observe that
a covariant channel $C$ that converts a state $\rho$ to another state 
$\sigma$ 
must necessarily convert $\overline{\rho}$ to $\overline{\sigma}$.
The channel $C$ is therefore only passive
if $F(\overline{\rho})\geq F(\overline{\sigma})$.   
This shows that asymmetry as well as covariant free energy are both 
non-increasing under passive covariant operations. We rephrase these observations
as a theorem:

\begin{Theorem}\label{MainVor}
The free energy of a quantum system with discrete energy levels can be
decomposed into
\[
F(\rho)=R(\rho) \,k_B T +F(\overline{\rho})\,,
\]
where
\[
R(\rho):=S(\overline{\rho})-S(\rho)
\]
is the Holevo information of the time orbit and $F(\overline{\rho})$ is
the free energy of the orbit average.
The terms $R(\rho)$ and $F(\overline{\rho})$ are both non-negative and
non-increasing with respect to time-covariant passive operations.   
\end{Theorem}

Theorem \ref{MainVor} can be generalized in two respects. First, we may have
an arbitrary group representation instead of the time evolution 
provided that it leaves the Hamiltonian invariant. Then the term
$tr(\rho H)$ is preserved by averaging, too. 
Second, we need not necessarily consider uniform averaging over the
whole group. Instead, we can define hierarchies of 
states, obtained by averaging more and more over the group, and
calculate free energy differences between more and less mixed states.
By this procedure, we obtain a splitting of free energy into many
independent terms.
We phrase this idea also
as a theorem:

\begin{Theorem}
Given  a quantum system with Hilbert space $\cH$ and Hamiltonian $H$.
Let $g\mapsto U_g$ with $g\in G$ 
 be the unitary representation of a group $G$ acting on
$\cH$ such that $[U_g,H]=0$. 
Let $\mu_1,\mu_2,\dots,\mu_n$ be an $n$-tuple of
probability measures 
on $G$ such that 
there exist measures $\nu_j$ on $G$ 
with
 $\mu_j* \nu_j=\mu_{j+1}$, i.e., $\mu_{j+1}$ is the convolution\footnote{The convolution product $\mu*\nu$ of measures $\mu,\nu$ on $G$ 
is here defined by the probability
distribution of
$
h \circ g 
$
if $g\in G$ and $h\in G$ are independently distributed according to 
$\mu$ and  $\nu$, respectively
(see \cite{HewittI}, and adapt Def. 19.8  to our setting).} 
of
$\mu_j$ with a third measure $\nu_j$. Let $\mu_1$ be the Dirac measure
on the identity. 

Let $A_\mu$ be the CP map 
given by the average
\[
A_\mu(\rho):= \int_G U_g \rho U_g^\dagger d\mu(g)\,.
\]

Then the free energy  $F(\rho)$  splits up into the $n$ terms
\[
F(\rho)=
\sum_{j=1}^{n} F_j(\rho)\,,
\]
with
\[
F_j(\rho):=F(A_{\mu_j}(\rho))-F(A_{\mu_{j+1}}(\rho))
\,\,\,\,\,j=1,\dots,n-1\,.
\]
and 
\[
F_n(\rho):=F(A_{\mu_n}(\rho))\,.
\]
All terms $F_j(\rho)$ for $j=1,\dots,n$ are non-negative and
non-increasing with respect
to passive covariant operations. 
\label{Th:Main}
\end{Theorem}

\begin{Proof}
$F_n(\rho)$ is clearly non-negative. To see that it is non-increasing
we observe 
\[
F\Big(A_{\mu_n}(C(\rho))\Big)=F\Big(C(A_{\mu_n}(\rho))\Big) 
\leq F(A_{\mu_n}(\rho))\,,
\]
where the last inequality is due to the monotonicity of usual 
free energy under passive operations. 
The terms $F_j(\rho)$ for $1\leq j\leq n-1$ are, up to the constant
$k_BT$, 
 given by the entropy difference
\begin{equation}\label{HolA}
S\Big(A_{\nu_j}(\rho_j)\Big)-S(\rho_j)\,,
\end{equation}
with
\[
\rho_j:=A_{\mu_j}(\rho)\,.
\]
This follows easily from $A_{\mu * \nu}=A_\nu \circ A_\mu$.
Expression~(\ref{HolA}) 
is for fixed $j$ the Holevo information \cite{NC} 
of the ensemble defined by 
the family of states $U_g \rho_j U_g^\dagger$ with $g\in G$ according to the 
probability measure $\nu_j$. It is therefore non-negative.
To show that it is non-increasing when applying $C$ we observe
that the covariance implies
\[
A_{\mu_j}(C(\rho))=C(\rho_j)\,,
\]
and $F_j(C(\rho))$ is therefore, up to the constant $k_B T$, 
the Holevo
information of the ensemble  
\[
U_g C(\rho_j) U_g^\dagger=C(U_g\rho_j \,U_g^\dagger), \,\,\,g\in G
\] 
according to the 
probability measure $\nu_j$.
Then monotonicity of
$F_j(\rho)$ with respect to $C$ follows 
from the monotonicity of Holevo information.
\end{Proof}

The advantage of Theorem \ref{Th:Main} compared to the preceding remarks 
is not only that it allows a splitting into more than two terms.
It is furthermore important that it allows a splitting for non-compact groups
since it does not refer to a uniform average over the whole group.

\section{Superadditive and subadditive components}

\label{Sec:Super}

We will now restrict our attention again to the splitting into
two free energy terms like in the beginning of Section \ref{Sec:Split}
and investigate how these quantities behave when systems are composed to
joint systems. Let us consider a two-level system with lower and upper 
state,  denoted by  $|0\rangle$ and $|1\rangle$, respectively. 
Let $E$ denote the energy gap between both levels. 
If $k_B T \geq E$ its equilibrium state is almost the maximally mixed state
$\gamma_\infty:= {\bf 1}_2/2$. 
The free energy of the state
\[
|+\rangle:=\frac{1}{\sqrt{2}}(|0\rangle+ |1\rangle)
\]
is given by
\[
F(|+\rangle \langle +|)=k_B T \,\Big(S(\overline{|+\rangle \langle +|})
-S(|+\rangle \langle +|)\Big) +F(\gamma_\infty)=\ln 2\,k_BT+F(\gamma_\infty) \,.
\]
One checks easily that $F(\gamma_\infty)$ is negligible compared
to the first term since $E\ll k_B T$ and the entropy difference
between $\gamma_\infty$ and $\gamma_T$ is small. 

To see the asymptotics of many copies of $|+\rangle \langle +|$ 
we observe that the entropy of the time average of 
$|+\rangle \langle +|^{\otimes n}$ is exactly the entropy 
of the binomial distribution $B_{n,1/2}$ with
\[
B_{n,1/2}(k):=\frac{1}{2^n}{n \choose k}\,. 
\]
Hence the  entropy of the average of $|+\rangle \langle +|^{\otimes n}$ 
increases only with $O(\ln n)$ since the measure is supported by 
 only $n$ different points. 
In other words, there are only
$n$ different eigenspaces of the joint Hamiltonian
\[
H=E \sum_j \sigma_z^{(j)}\,,
\]
where $\sigma_z^{(j)}$ is the Pauli matrix $\sigma_z$ acting
on qubit $j$. 
For the same reasons, the 
covariant free energy  of $|+\rangle \langle +|^{\otimes n}$ 
can be bounded from below by $n \ln 2 -O(\ln n)$.
Hence the asymmetry part of the free energy is for large $n$ more and more
a negligible fraction of the total free energy. 
 Similar arguments apply to the general situation.

This does not mean, however, that the splitting is completely irrelevant
for large particle numbers. Instead, the results show that 
every process that increases the covariant free energy of some of the
 particles 
requires interactions between them. Hence the amount of increase of 
covariant free energy can bound the number of interacting particles 
from below.
Therefore the results imply statements on the {\it complexity} of the 
considered process. This kind of complexity issues are related to the
questions discussed in \cite{HeatEngines} where we have discussed 
the complexity  of molecular heat engines. The key observation was that
the additional constraints for energy conversion processes that arise in
simple quantum systems imply statements on the complexity of
energy conversion processes in macroscopic ensembles of particles.

For the sake of completeness we sketch the proof of the 
superadditivity of covariant
free energy. 
Then the subadditivity of the asymmetry term follows 
because total free energy is additive. 
The essential observation is that for two states $\rho$ and $\sigma$ 
with time evolution $\alpha_t$ and $\beta_t$, respectively, the entropy
of the joint time average $\overline{\rho\otimes \sigma}$ 
cannot be greater than the entropy of the tensor product
of the averages, i.e., $\overline{\rho}\otimes \overline{\sigma}$.
This is because  the latter state
can be obtained from  the former
by averaging over all possible 
{\it relative} time translations $\alpha_t \otimes\beta_{-t}$. 
We have therefore 
\[
F(\overline{\rho\otimes \sigma})\geq F(\overline{\rho}\otimes \overline{\sigma})=F(\overline{\rho})+F(\overline{\sigma})\,.
\]

\section{Applications}

\label{Sec:Appl}

Remarkably, the results above have also implications for situations
where a reference frame is available 
since every passive covariant operation that decreases one term $F_j(\rho)$
necessarily decreases the total free energy since it is unable to convert
one kind of free energy into the other. 
This can be used to derive
lower bounds on the loss of free energy of a physical signal 
like an electrical pulse or a light pulse 
when it passes
a device such that it degrades the time accuracy of the pulse.
We will explain this idea using a system  with discrete energy levels
as a toy model for the physical signal.

Consider for instance a system which has some integer values $n_1,n_2,\dots,
n_d$ 
as energy spectrum. Due to the periodicity, we can restrict 
its group of time translations to $G:=SU(1)$. 
Parameterizing $G$ by the interval $[0,2\pi]$ we obtain
 the unitary 
representation  
\[
t \mapsto U_t:=diag(e^{-it n_1},e^{-it n_2},\dots, e^{-it n_d})\,.
\]  
Now we consider an input 
state $\rho$ which is perfectly distinguishable from
its time evolved state $\alpha_s(\rho)$ for some $s\in \R$ in the sense that
\[
tr(\rho \,\alpha_s(\rho))=0\,.
\]
Let $C$ be a passive 
time covariant operation that corrupts the timing information
of $\rho$ in the sense that the
corresponding output density matrices $C(\rho)$ and
$C(\alpha_s(\rho))=\alpha_s(C(\rho))$ are not perfectly distinguishable.
Then we can use the results of Section \ref{Sec:Split} to bound the 
free energy loss of the channel from below as follows. 
Define a measure on $SU(1)$ by 
\[
\mu:=\frac{1}{2}(\delta_0 +\delta_s)\,,
\]
where $\delta_0$ and $\delta_s$ denote the Dirac measures 
at the time  instants $t=0$ and  $t=s$, respectively.
By applying Theorem \ref{Th:Main} using the measures $\mu_1:=\delta_0$ and
$\mu_2:=\mu$, 
the free energy of the input splits up into 
\[
F(\rho)=k_B T\, \Big(S(A_\mu(\rho))-S(\rho)\Big) 
+F(A_\mu(\rho))=k_BT \,\ln 2 +F(A_\mu(\rho))\,,
\]
and for the output into 
\[
F(C(\rho))=k_B T \, \Big(S(A_\mu(C(\rho)))-
S(C(\rho))\Big)+F\Big(A_\mu(C(\rho))\Big)\,.
\]
The fact that the output states are not perfectly distinguishable 
is equivalent to the statement that the Holevo information of 
an ensemble that consists of the states $C(\rho)$ and $C(\alpha_s(\rho))$ 
with probability $1/2$ each is strictly less than $\ln 2$, i.e.,  
$c:=S(A_\mu(C(\rho)))-S(C(\rho))<\ln 2$. 
Then the loss of free energy satisfies
\[
F(\rho)-F(C(\rho))\geq (\ln 2- c)\, k_B T\,.
\]
The intuitive content of this statement is that it provides lower bounds
on  
the free energy loss caused by devices that corrupt the time accuracy of
the input signal by generating output signals with stochastic time delay. 

To consider a more concrete physical  situation, 
assume some electrical, acoustical, or optical signal enters a passive device
whose input-output behavior is described by the time 
covariant map $C$.
The time covariance reflects only
the fact that 
the state of the physical device is stationary 
before the signal enters into 
it (see \cite{TimeCovariant} for details). 
Assume that the channel
converts
some 
input signal 
that can be perfectly distinguished from its
time evolved copy that is defined by a time shift  $\Delta t$
in the sense that the quantum states $\rho$ and 
$\alpha_{\Delta t}(\rho)$ 
are mutually orthogonal density operators. 
Assume that the channel generates an unknown time delay such that the
output density operators $C(\rho)$ and $C(\alpha_{\Delta t}(\rho))$ are 
not perfectly distinguishable.

\begin{figure}
\centerline{
\epsfbox[0 0 364 78]{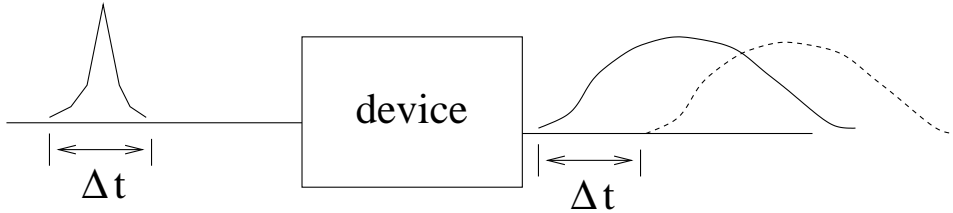}
}
\caption{{\small Symbolic drawing of input and output signals. The curves do not
necessarily  have direct physical meaning. Their widths  only 
symbolize the time scale of distinguishability. The curves could, however,
have a direct meaning in the following situation. 
Consider a pulse which has on the considered time scale a well-defined
time of arrival (since the quantum uncertainty \cite{AhaArrival} 
may  be only relevant on a
much smaller scale). Assume that  
the time of arrival is subjected to stochastic fluctuations   
such that the curves indicate the probability distribution of the 
time of arrival. Then an increase of these fluctuations leads necessarily to a loss of free energy according to our results.}} 
\end{figure}

Then the fluctuation of the time delay leads necessarily 
to a loss of free energy. One may argue that
these fluctuations would obviously lead to a loss of free energy
of the signal since they increase the entropy of the state.
However, a priori it is not clear whether the channel could change the 
probability distribution of energy values such  that the loss of free 
energy caused by an increase of entropy 
is compensated by an increase of the inner energy $tr(\rho H)$. 
The statement that no covariant channel can do such a compensation is
the key statement of this article.

The signal above is assumed to pass the device only once. 
By applying our results to such a situation with aperiodic dynamics 
we have actually ignored the
fact that it refers necessarily to continuous spectrum. Otherwise
the free evolution of the signal would be quasiperiodic. 
We may remove this by letting it oscillate between distant mirrors
for obtaining discrete energy spectrum. The problem with
the aperiodic limit is anyway that it  refers 
 to an infinite amount of free energy. This can be seen as follows. 
Given a density operator $\rho$ such that for some $t\in \R$ 
all states 
$\alpha_{nt}(\rho)$ for $n\in \Z$ are perfectly distinguishable from $\rho$.  
Then one can choose an arbitrary probability measure on $\Z$ by $(p_j)$ 
with $\sum_j p_j=1$ and observe
\[
F(\rho) -F\Big(\sum_j p_j \alpha_{tj} (\rho)\Big) =k_B T\, S(p)\,,
\]
where $S(p)$ denotes the Shannon entropy of $p$.  By choosing measures $p$ 
with diverging entropy one can show that the free
and recalling that $F$ is always non-negative, the statement follows.
However,
even though the free energy diverges in the aperiodic limit our statement
on the free energy {\it loss}  still makes sense since the absolute value
of the free energy is irrelevant in this context.

\vspace{0.5cm}

This work was funded by the Landesstiftung 
Baden-W\"{u}rttemberg, project AZ1.1422.01.

%\bibliographystyle{unsrt}

%\bibliography{/mnt/boleyn/users1/janzing/EigenePapers/literatur}

\begin{thebibliography}{10}

\bibitem{Szilard}
L.~Szilard.
\newblock {\"{U}ber die Entropieverminderung in einem thermodynamischen System
  bei Eingriffen intelligenter Wesen}.
\newblock {\em Z. Phys}, pages 840--856, 1929.

\bibitem{JWZGB}
D.~Janzing, P.~Wocjan, R.~Zeier, R.~Geiss, and Th. Beth.
\newblock {Thermodynamic cost of reliability and low temperatures : Tightening
  Landauer's principle and the Second Law}.
\newblock {\em Int. Jour. Theor. Phys.}, 39(12):2217--2753, 2000.

\bibitem{Ja00}
D.~Janzing.
\newblock A quasi-order of resources as a new concept for a thermodynamic
  theory of quantum state preparation.
\newblock In {\em Sciences of the interface}, {T\"{u}bingen}, 2000. Genista
  Verlag.

\bibitem{Landauer:61}
R.~Landauer.
\newblock Irreversibility and heat generation in the computing process.
\newblock {\em IBM J. Res. Develop.}, 5:183--191, 1961.

\bibitem{BennettThermoReview}
C.~Bennett.
\newblock The thermodynamics of computation -- a review.
\newblock {\em Int. J. Theor. Phys.}, 21:905--940, 1982.

\bibitem{Callen}
H.~Callen.
\newblock {\em Thermodynamics}.
\newblock J. Wiley and Sons, New York, 1960.

\bibitem{AllahverdyanGibbs}
A.~Allahverdyan and T.~Nieuwenhuizen.
\newblock {Resolution of the Gibbs paradox via quantum thermodynamics}.
\newblock {\em http://xxx.lanl.gov/abs/quant-ph/0507145}.

\bibitem{OhyaPetz}
M.~Ohya and D.~Petz.
\newblock {\em Quantum entropy and its use}.
\newblock Springer Verlag, 1993.

\bibitem{AllahverdyanNano}
A.~Allahverdyan, R.~Balian, and T.~Nieuwenhuizen.
\newblock Quantum thermodynamics: thermodynamics at the nanoscale.
\newline {\em http://xxx.lanl.gov/abs/cond-mat/0402387}.

\bibitem{Allahverdyan}
A.~Allahverdyan, R.~Balian, and T.~Nieuwenhuizen.
\newblock Maximal work extraction from quantum systems.
\newblock {\em Europhys. Lett}, 67:565, 2004.

\bibitem{clock}
D.~Janzing and T.~Beth.
\newblock Quasi-order of clocks and their synchronism and quantum bounds for
  copying timing information.
\newblock {\em IEEE Trans. Inform. Theor.}, 49(1):230--240, 2003.

\bibitem{RefEfficient}
G.~Chiribella, G.~D'Ariano, P.~Perinotti, and M.~Sacchi.
\newblock Efficient use of quantum resources for the transmission of a
  reference frame.
\newblock {\em http://xxx.lanl.gov/abs/quant-ph/0405095v2}.

\bibitem{GroverRudolph}
T.~Rudolph and L.~Grover.
\newblock On the communication complexity of establishing a shared reference
  frame.
\newblock {\em http://xxx.lanl.gov/abs/quant-ph/0306017}.

\bibitem{Decoherence-full}
S.~Bartlett, T.~Rudolph, and W.~Spekkens.
\newblock Decoherence-full subsystems and the cryptographic power of a private
  reference frame.
\newblock {\em http://xxx.lanl.gov/abs/quant-ph/04031v2}.

\bibitem{SynchrEntropy}
D.~Janzing and T.~Beth.
\newblock Synchronizing quantum clocks with classical one-way communication:
  Bounds on the generated entropy.
\newblock {\em http://xxx.lanl.gov/abs/quant-ph/0306023v1}.

\bibitem{GroupCovariantThermo}
J.~Vaccaro, F.~Anselmi, H.~Wiseman, and K.~Jacobs.
\newblock Complementarity between extractable mechanical work, accessible
  entanglement, and ability to act as a reference frame, under arbitrary
  superselection rules.
\newblock {\em http://xxx.lanl.gov/abs/quant-ph/0501121}.

\bibitem{NC}
M.~Nielsen and I.~Chuang.
\newblock {\em Quantum Computation and Quantum Information}.
\newblock Cambridge University Press, 2000.

\bibitem{TimeCovariant}
D.~Janzing.
\newblock Decomposition of time-covariant operations on quantum systems with
  continuous and/or discrete energy spectrum.
\newblock {\em http://xxx.lanl.gov/abs/quant-ph/0407144, to appear in Journ.
  Math. Phys.}

\bibitem{PoulinRelation}
D.~Poulin.
\newblock Toy model for a relational formulation of quantum theory.
\newblock {\em http://xxx.lanl.gov/abs/quant-ph/0505081}.

\bibitem{oppenheim}
M.~Horodecki, P.~Horodecki, and J.~Oppenheim.
\newblock Reversible transformations from pure to mixed states, and the unique
  measure of information.
\newblock {\em Phys. Rev. A}, 67:062104, 2003.

\bibitem{LindOp}
G.~Lindblad.
\newblock Completely positive maps and entropy inequalities.
\newblock {\em Comm. Math. Phys.}, 40:147--151, 1975.

\bibitem{HewittI}
E.~Hewitt and K.~Ross.
\newblock {\em Abstract Harmonic Analysis}, volume~I.
\newblock Springer, 1963.

\bibitem{HeatEngines}
D.~Janzing.
\newblock On the computational power of molecular heat engines.
\newblock {\em To appear in J. Stat. Phys., see also
  http://xxx.lanl.gov/abs/quant-ph/0502019}.

\bibitem{AhaArrival}
Y.~Aharonov, J.~Oppenheim, S.~Popescu, B.~Reznik, and W.~Unruh.
\newblock Measurement of time-of-arrival in quantum mechanics.
\newblock {\em Phys. Rev. A}, 57:4130, 1998.

\end{thebibliography}

\end{document}